\documentclass[prl,reprint,superscriptaddress,preprintnumbers]{revtex4-1}
\pdfoutput=1
\usepackage{amsfonts, amssymb, amsmath}
\usepackage{mathrsfs}
\usepackage{hyperref}
\usepackage[T1]{fontenc}

\def\im{{\rm i}} 

\usepackage{color}
\definecolor{dark-gray}{gray}{0.20}
\definecolor{gray}{gray}{0.30}
\definecolor{light-gray}{gray}{0.80}
\definecolor{dark-red}{rgb}{0.7,0,0}
\definecolor{dark-green}{rgb}{0.1,0.4,0}
\definecolor{dark-blue}{rgb}{0.3,0.3,0.7}
\definecolor{light-blue}{rgb}{0.8,0.8,1}
\definecolor{blue}{rgb}{0,0,1}
\definecolor{red}{rgb}{1,0,0}
\definecolor{green}{rgb}{0,1,0}

\usepackage{hyperref}
\hypersetup{
	colorlinks=true,
	linkcolor=dark-blue,
	citecolor=dark-red,
	urlcolor=dark-blue,
	linktoc=page
}

\def\coeff#1#2{\relax{\textstyle {#1 \over #2}}\displaystyle}

\def\RR{\mathbb{R}}

\def\SO{{\rm SO}}

\def\SU{{\rm SU}}

\newcommand{\dd}{\mathrm{d}}

\newcommand{\e}{\mathrm{e}}
\newcommand{\w}{\wedge}

\newcommand{\be}{\begin{equation}}
\newcommand{\ee}{\end{equation}}
\newcommand{\bea}{\begin{eqnarray}}
\newcommand{\eea}{\end{eqnarray}}

\newcommand{\f}[2]{\frac{#1}{#2}}

\newcommand{\p}[1]{\phantom{#1}}
\newcommand{\vol}{\text{vol}}

\newcommand{\U}{\text{U}}

\def\vol {\mathrm{vol}}

\begin{document}

\title{The Holographic Dual of the $\Omega$-background}

\date{\today}

\author{Nikolay Bobev}

\affiliation{Instituut voor Theoretische Fysica, KU Leuven, Celestijnenlaan 200D, B-3001 Leuven, Belgium}

\author{Fri\dh rik Freyr Gautason}

\affiliation{Instituut voor Theoretische Fysica, KU Leuven, Celestijnenlaan 200D, B-3001 Leuven, Belgium}

\author{Kiril Hristov}

\affiliation{INRNE, Bulgarian Academy of Sciences, Tsarigradsko Chaussee 72, 1784 Sofia, Bulgaria}

\begin{abstract}
\noindent We find an explicit supergravity background dual to the $\Omega$-deformation of a four-dimensional $\mathcal{N}=2$ SCFT on $\mathbb{R}^4$. The solution can be constructed in the five-dimensional ${\cal N}=4^+$ gauged supergravity and has a nontrivial self-dual 2-form. When uplifted to type IIB supergravity the background is a deformation of AdS$_5\times S^5$ which preserves 16 supercharges. We also discuss generalizations of this solution corresponding to turning on a vacuum expectation value for a scalar operator in the dual SCFT.

\end{abstract}


\maketitle

\section{Introduction}\label{sec:intro}
%
The $\Omega$-deformation was introduced by Nekrasov in \cite{Nekrasov:2002qd} as a tool to calculate the path integral of four-dimensional $\mathcal{N}=2$ gauge theories via supersymmetric localization. The deformation can be thought of as a supersymmetric modification of the gauge theory Lagrangian on $\mathbb{R}^{4}$ by two real parameters, $\epsilon_1$ and $\epsilon_2$. The two deformation parameters are associated with two Killing vectors on $\mathbb{R}^{4}$ and are used to define an appropriate equivariant cohomology. This ultimately leads to a rigorous evaluation of the Nekrasov partition function of the field theory, $Z(\epsilon_1,\epsilon_2,\ldots)$. Here the dots stand for possible dependence of the path integral on various deformation parameters, like Coulomb branch vacuum expectation values (vevs) and superpotential mass terms, compatible with supersymmetry. Expanding this partition function in the limit $\epsilon_{1,2} \to 0$ leads to an exact evaluation of various physical quantities in the undeformed theory on $\mathbb{R}^4$. For example, the leading term in this expansion is the Seiberg-Witten prepotential on the Coulomb branch of the theory \cite{Seiberg:1994rs,Seiberg:1994aj}. One can further generalize this construction by introducing the $\Omega$-deformation for more general four-manifolds which posses a Killing vector \cite{Nekrasov:2003rj}. This can be thought of as an extension of the Donaldson-Witten twist of four-dimensional $\mathcal{N}=2$ QFTs \cite{Witten:1988ze,Donaldson:1990kn}. 

The $\Omega$-deformation and the corresponding Nekrasov partition function find many applications in the physics (and mathematics) of four-dimensional $\mathcal{N}=2$ QFTs. For example, they are instrumental in the AGT \cite{Alday:2009aq} correspondence and the Nekrasov-Shatashvili correspondence \cite{Nekrasov:2009rc}. In addition the Nekrasov partition function can be thought of as a ``building block'' for a plethora of exact results for supersymmetric QFTs on compact curved manifolds, see \cite{Pestun:2016zxk} for a review.

Given the importance of the $\Omega$-deformation in many supersymmetric localization calculations and the fruitful interplay between localization and holography for four-dimensional $\mathcal{N}=2$ QFTs, see for example \cite{Buchel:2013id,Bobev:2013cja}, it is natural to ask what is the holographic dual description of the $\Omega$-deformation. Our goal here is to answer this question for SCFTs with weakly coupled supergravity duals.

Rather than viewing the $\Omega$-deformation as a modification of the path integral of the supersymmetric QFT one can consider it as a background of rigid four-dimensional $\mathcal{N}=2$ supergravity, see for example \cite{Hama:2012bg,Klare:2013dka}. We will call this the $\Omega$-background.  The $\Omega$-background can be defined for any four-dimensional $\mathcal{N}=2$ QFT but from the holographic perspective it is easiest to describe it for superconformal theories in which all dimensionful couplings and vevs are turned off. In addition the supergravity approximation requires that we work in the planar limit and at large 't Hooft coupling. In this context the holographic dual of the $\Omega$-background is a simple modification of the well-known vacuum AdS$_5$ solution of five-dimensional $\mathcal{N}=4^{+}$ gauged supergravity \cite{Romans:1985ps}. We present this solution explicitly and show how to embed it in type IIB supergravity where it describes holographically the $\Omega$-background for the $\mathcal{N}=4$ SYM theory.

\section{The $\Omega$-background}\label{sec:fieldtheory}
%

As discussed in \cite{Festuccia:2011ws} there is a systematic way to classify non-trivial backgrounds which preserve some of the supersymmetry of a given $d$-dimensional QFTs. This formalism is based on using off-shell supergravity in $d$ dimensions. For superconformal theories one needs to use off-shell superconformal gravity to address the same question, see \cite{Klare:2012gn}. For four-dimensional $\mathcal{N}=2$ SCFTs this analysis was initiated in \cite{Klare:2013dka} and here we summarize the salient features of their results relevant for the $\Omega$-background.

The four-dimensional $\mathcal{N}=2$ Weyl multiplet of Euclidean conformal supergravity consists of the metric $g^{(4)}_{\mu\nu}$, two gravitini $\psi_\mu^i$, two real 2-forms $T^{\pm}_{\mu\nu}$, $\SO(1,1)$ and $\SU(2)$ gauge fields, $\mathcal{A}^0_\mu$ and $\mathcal{A}^{ij}_\mu$, two dilatini $\chi^i$, and a real scalar field $\tilde{d}$, see \cite{Toine} for a review. In the rigid supergravity limit these fields capture the couplings to all operators in the energy-momentum multiplet of an $\mathcal{N}=2$ SCFT. A  supersymmetric bosonic background for the SCFT is fully specified by the fields in the Weyl multiplet as well as a conformal Killing spinor parameter which encodes the preserved supersymmetry. It was shown in \cite{Klare:2013dka}, in agreement with \cite{Hama:2012bg}, that the $\Omega$-background is described in this formalism by the following bosonic fields
\begin{align}\label{eq:Omega-def}
\begin{split}
	{\rm d} s_4^2 &= \dd x_1^2+\dd x_2^2+\dd x_3^2+\dd x_4^2\, , \\ 
T^- &=\dd b\,,\qquad b = 2\beta (x_{[2}\dd x_{1]}+x_{[4}\dd x_{3]})\,.
\end{split}
\end{align} 
The rest of the fields in the Weyl multiplet vanish. The parameters specifying the conformal Killing spinors preserved by this background are
\begin{equation}\label{eq:4dspinors}
\begin{split}
\zeta^{-} &= \zeta_{0}^{-}-\coeff{\im}{2}b_m\gamma^{m}\zeta_{0}^{+}+x^{m}\gamma_{m}\eta_0^{+}\,,\\
\zeta^{+} &=\zeta_{0}^{+}\,, \qquad \eta^{\pm} = \coeff{1}{4}\gamma^{m}\partial_{m}\zeta^{\mp}\,.
\end{split}
\end{equation}
Here $\gamma_m$ are Dirac $\gamma$-matrices, $\zeta_{0}^{\pm}$ and $\eta_{0}^+$ are constant spinors with a definite chirality, $\gamma_{1234}\zeta_{0}^{\pm} = \pm \zeta_{0}^{\pm}$, $\gamma_{1234}\eta_{0}^{+} = + \eta_{0}^{+}$, and $b$ is the 1-form on $\mathbb{R}^4$ given in \eqref{eq:Omega-def}. Note that we have twelve independent real components in the spinor parameters in \eqref{eq:4dspinors} since $\eta_{0}^{-}=0$. From the perspective of the four-dimensional $\mathcal{N}=2$ SCFT in flat space $\zeta_{0}^{+}$ and $\zeta_{0}^{-}$ generate the eight $Q$ supercharges and $\eta_0^{+}$ generate four of the $S$ supercharges \footnote{Note that in \cite{Hama:2012bg,Klare:2013dka} only four of the supercharges were explicitly considered. These four can be recovered from the full expression in \eqref{eq:4dspinors} by setting $\zeta_0^-$ to zero and $\eta_0^+$ proportional to $\gamma_{12}\zeta_0^+$.}. The $\Omega$-background in \eqref{eq:Omega-def} thus breaks the $S$ supercharges associated to $\eta_{0}^{-}$

There is an analogous supersymmetric background to the one in \eqref{eq:Omega-def} with a non-vanishing anti self-dual 2-form $T^+$ instead of the self-dual $T^-$. The preserved spinors for it are the same as in \eqref{eq:4dspinors} but with $\eta_{0}^{+}$ replaced by $\eta_0^{-}$. Note that the bosonic background in \eqref{eq:Omega-def} depends only on a single parameter $\beta$ which should be thought of as a linear combination of the $\Omega$-deformation parameters $\epsilon_{1,2}$. In \cite{Hama:2012bg,Klare:2013dka} the parameter $\beta$ was identified with the linear combination $\epsilon_1+\epsilon_2$. We will keep using the parameter $\beta$ to denote the specific deformation studied in this paper.

The background in \eqref{eq:Omega-def} describes a deformation of the SCFT involving only operators in the energy-momentum tensor multiplet. For Lagrangian SCFTs the 2-form coupling $T^{-}$ in \eqref{eq:Omega-def} turns on a deformation by a dimension 3 operator of the schematic form $\text{Tr}[\Phi F^{+}_{\mu\nu}+\bar{\xi}\gamma_{\mu\nu}^{+}\xi]$. Here $\Phi$ is the complex scalar in a vector multiplet with gauge field $F_{\mu\nu}$, $\xi$ are the fermions in a hyper multiplet, the trace is over gauge indices, and the $+$ superscript denotes the self-dual part of a 2-form. This is the minimal deformation that the $\mathcal{N}=2$ SCFT is subjected to in the $\Omega$-background. The conformal symmetry and the $\SO(1,1)$ R-symmetry are broken by the deformation but a linear combination of the dilatation operator and the generator of $\SO(1,1)$ is preserved. To describe the coupling of other operators in the SCFT in the presence of the $\Omega$-deformation one should study the more general rigid supergravity setup in which the Weyl multiplet is coupled to background vector and hyper multiplets. This should also allow for the study of $\Omega$-deformations of $\mathcal{N}=2$ SCFTs for which the two deformation parameters $\epsilon_{1,2}$ are independent.

As discussed in \cite{Klare:2012gn} every supersymmetric background of conformal supergravity in $d$ dimensions can also be viewed as a supersymmetric asymptotically locally AdS boundary condition for an appropriate gauged supergravity in $d+1$ dimensions. This vantage point makes it clear that to construct the holographic dual to an SCFT placed on this $d$-dimensional background one has to solve the full set of supersymmetry variations and equations of motion of the gauged supergravity theory. This is in general a non-trivial technical task which should be addressed on a case by case basis and may not always lead to regular bulk solutions. We now show how to implement this program for the $\Omega$-background in \eqref{eq:Omega-def}.

\section{${\cal N} = 4^+$ supergravity}\label{sec:sugra}
%
To construct the holographic dual to the four-dimensional rigid supergravity background above we use the ${\cal N} = 4^+$ gauged supergravity theory in five-dimensions \cite{Romans:1985ps}. The bosonic dynamical fields in this theory are the metric, $g^{(5)}_{\mu\nu}$, a scalar field, an $\SU(2)\times \U(1)$ gauge field and a pair of 2-forms. There are also four gravitini as well as four gaugini. These fields comprise the bulk counterpart of the four-dimensional $\mathcal{N}=2$ Weyl multiplet discussed above. The background of interest is captured by an Euclidean version of the ${\cal N} = 4^+$ gauged supergravity theory. This analytic continuation changes the Abelian gauge group from $\U(1)$ to $\SO(1,1)$, see \cite{BenettiGenolini:2017zmu}. The bosonic Lagrangian of the Euclidean ${\cal N} = 4^+$ gauged supergravity is \footnote{We use the conventions of \cite{BenettiGenolini:2017zmu} but reinstate the explicit gauge coupling constant $g$ and rescale the 2-form and the gauge fields by a factor of 2. See \cite{BBGH,BGH} for more details.} 
\begin{align}\label{eq:Romanslagrangian}
\begin{split}
{\cal L} =&  \sqrt{g^{(5)}}\Big[R-\frac12 |\dd\lambda|^2 + 2X^4|f|^2  - V(\lambda) \\ -& X^{-2}\left(\text{tr} |F|^2 +B^+\cdot B^- \right)\Big] \\
 -& \frac{1}{g}\left(B^+\wedge H^- -B^-\wedge H^+\right)  - 2\, \text{tr} (F\w F) \w a~,
\end{split}
\end{align}
where $g$ is the gauge coupling, $X= \e^{-\lambda/\sqrt{6}}$ is the scalar, $a$ is the $\rm SO(1,1)$ gauge field with field strength $f=\dd a$, $A^i$ is the $\SU(2)$ gauge field with field strength $F^i =\dd A^i + g\epsilon^i_{\p{i}jk}A^j\w A^k$, and $B^\pm$ is a 2-form doublet  which is charged under the $\SO(1,1)$ gauge field:
\begin{equation}
H^\pm = \dd B^\pm \mp  g\,a\w B^\pm~.
\end{equation}
The 2-forms $B^\pm$ are massive and obey an odd-dimensional self-duality condition
\begin{equation}\label{eq:oddselfudality}
H^\pm \mp \f{g}{2} X^{-2} \star_5B^\pm = 0~.
\end{equation}
The scalar potential can be derived from a real superpotential $W \equiv g(2X+X^{-2}),$ and takes the form
\begin{equation}\label{DWPot}
V(\lambda) = \coeff{1}{2} (\partial_\lambda W)^2 -\coeff{1}{3} W^2= - g^2(X^2 + 2X^{-1})~.
\end{equation}
The supersymmetry variations of the fermions read
\begin{align}
\begin{split}
\delta \psi_\mu &= \left(D_\mu - \coeff{1}{12}\gamma_\mu W~ \hat\sigma_{3} + \coeff{\im}{12}(\gamma_{\mu}^{\p{\mu}{\nu\rho}} - 4\delta_\mu^{\nu}\gamma^\rho)h_{\nu\rho} \right)\epsilon\,,\\
\delta \chi &= -\coeff{\im}{2\sqrt{2}}\left(\gamma^\mu\partial_\mu \lambda + \partial_\lambda W~ \hat\sigma_{3} + \im\gamma^{\mu\nu}\partial_\lambda h_{\mu\nu} \right)\epsilon\,,\label{bpsvars}
\end{split}
\end{align}
where we have defined
\begin{align}
h_{\mu\nu} = \frac{1}{X}\left( F_{\mu\nu}^i \hat\sigma_3 \sigma_i + B_{\mu\nu}^+\hat{\sigma}_{-} + B_{\mu\nu}^-\hat{\sigma}_{+} \right) - \im X^2 f_{\mu\nu}\,,
\end{align}
with $\hat{\sigma}_\pm = (\hat\sigma_1\pm \im\hat\sigma_2)/2$. Here $\gamma_\mu$ denote spacetime $\gamma$-matrices whereas $\sigma_i$ and $\hat\sigma_i$, for $i=1,2,3$, denote two commuting sets of Pauli matrices. The parameter $\epsilon$ represents a pair of spinors with a definite charge under $\SO(1,1)$ in the doublet of $\SU(2)$. 

The equations of motion for this five-dimensional theory can be readily derived from the Lagrangian in \eqref{eq:Romanslagrangian} and are explicitly given in \cite{BenettiGenolini:2017zmu,BBGH,BGH}.
%
\section{The supergravity solution}\label{sec:bulksol}
%
For the solution of interest most bosonic fields of the supergravity theory vanish
\begin{equation}
\lambda=0\,,\quad a = 0\,, \quad A^i = 0\,, \quad B^+ = 0\,.
\end{equation}
The metric is that of Euclidean AdS$_5$ in Poincare coordinates,
\begin{equation}\label{eq:AdS5metric}
\dd s_5^2 = \frac{L^2}{z^2} (\dd z^2 + \dd s^2_4)\, ,
\end{equation}
with ${\rm d} s_4^2$ the flat metric on $\mathbb{R}^4$ in \eqref{eq:Omega-def}. The length scale of AdS is fixed in terms of the coupling constant $g$ as $L=2/g$. The only other non-vanishing field is
\begin{equation}\label{eq:Bminsol}
B^- = - \frac{L}{z} \beta\  ({\rm d} x_1 \wedge {\rm d} x_2 + {\rm d} x_3 \wedge {\rm d} x_4)\, ,
\end{equation}
where $\beta$ is a real parameter which is the bulk counterpart of the parameter in \eqref{eq:Omega-def}. It is not hard to check that this simple bosonic background solves all equations of motion of the supergravity theory. The boundary of  AdS$_5$ is located at $z=0$ and the 2-form in \eqref{eq:Bminsol} diverges there as $1/z$. This is the appropriate asymptotic behavior to source the 2-form operator of dimension $\Delta = 3$ dual to $B^{-}$ in the $\mathcal{N}=2$ SCFT. This source is given precisely by the 2-form $T^{-}$ in \eqref{eq:Omega-def}. 

This solution can be generalized by giving the scalar field $X$ a non-trivial profile. This extension also allows for an analytic solution with metric
\begin{equation}\label{eq:metCoulomb}
{\rm d} s_5^2 = \frac{L^2}{X^2z^2} (\dd z^2 + X^3\dd s^2_4)\, ,
\end{equation}
and scalar and 2-form given by
\begin{equation}\label{eq:XBCoulomb}
\begin{split}
X^3 &= 1+ w z^2\,,\\
B^- &= - \frac{LX^{3/2}}{z} \beta\  ({\rm d} x_1 \wedge {\rm d} x_2 + {\rm d} x_3 \wedge {\rm d} x_4)\,.
\end{split}
\end{equation}
The metric is no longer that of AdS$_5$ and the solution is therefore not dual to the conformal vacuum of the $\Omega$-deformed ${\cal N}=2$ SCFT. Instead, the background is dual to a non-trivial supersymmetric vacuum state in which conformal symmetry is broken. This breaking is controlled by the integration constant $w$ which is proportional to the vacuum expectation value of the dimension $\Delta=2$ SCFT operator dual to the scalar $X$. The range of the coordinate $z$ is $z\in (0,\infty)$ for $w>0$, and $z\in (0,\sqrt{-w}]$ for $w<0$ and the solution has a naked singularity at the upper end of this interval.  Applying the criterion proposed in \cite{Gubser:2000nd} one finds that the singularity is acceptable in string theory. This is due to the fact that the potential in \eqref{DWPot} is manifestly bounded above.

The vanishing of the dilatino variation in \eqref{bpsvars} imposes
\begin{equation}\label{eq:dilproj}
(1-\gamma_{1234})(1-\hat{\sigma}_3)\epsilon=0\,.
\end{equation}
The gravitino variation in \eqref{bpsvars} is then solved by the following explicit spinor
\begin{equation}\label{eq:5dspinor}
\begin{split}
\epsilon &=  z^{-1/2} \varepsilon_0^{-}{}_{+}+(z^{1/2}+z^{-1/2}x^{m}\gamma_{m})\varepsilon_0^{+}{}_{+}\\
&\qquad\qquad\qquad\qquad+z^{-1/2}\left(1-\coeff{\im}{2}\, b_{m}\gamma^{m}\hat{\sigma}_{+}\right)\varepsilon_0^{+}{}_{-}\,.
\end{split}
\end{equation}
Here $\varepsilon_0$ are an $\SU(2)$ doublet of constant spinors with definite chirality and $\SO(1,1)$ eigenvalues, namely $\gamma_{1234}\varepsilon_0^{\pm}=\pm \varepsilon_0^{\pm}$ and $\hat{\sigma}_{3}\varepsilon_0\phantom{}_{\pm} = \pm\varepsilon_0\phantom{}_{\pm}$, and $b^{m}$ is defined in \eqref{eq:Omega-def}. The constant spinor parameter $\varepsilon_0^{-}{}_{-}$ vanishes due to \eqref{eq:dilproj} and thus the background in (\ref{eq:AdS5metric},\,\ref{eq:Bminsol}) preserves 12 real supercharges. These correspond to the 12 real supercharges discussed below \eqref{eq:4dspinors} in the rigid supergravity context. The more general background in (\ref{eq:metCoulomb},\,\ref{eq:XBCoulomb}) with $w\neq0$ preserves 8 real supercharges. The explicit spinor is a generalization of the one in \eqref{eq:5dspinor} with $\varepsilon_0^{+}{}_{+}=0$ and a different dependence on the radial coordinate $z$.

The background in \eqref{eq:AdS5metric} and \eqref{eq:Bminsol} preserves only  part of the bosonic generators of the superconformal algebra. All four special conformal generators are broken and the rotation group is broken from $\SO(4)$ to $\SU(2)\times \U(1)$. The four momentum generators as well as the $\SU(2)$ R-symmetry are preserved. Only a linear combination of the dilatation generator $D$ and the $\SO(1,1)$ generator $r$ is preserved. More details on this supersymmetric algebra will be presented in \cite{BGH}.

In analogy to the four-dimensional field theory discussion, in \eqref{eq:XBCoulomb} one can turn off the 2-form $B^{-}$ and turn on $B^{+}$ instead. The only modification is that $B^{+}$ should be anti-self-dual on the boundary, i.e. with $({\rm d} x_1 \wedge {\rm d} x_2 - {\rm d} x_3 \wedge {\rm d} x_4)$ on the right hand side of \eqref{eq:XBCoulomb}.

The supergravity solution in (\ref{eq:AdS5metric},\,\ref{eq:Bminsol}) describes the $\Omega$-deformation of all four-dimensional $\mathcal{N}=2$ SCFTs with a weakly coupled gravity dual and thus should capture universal properties of these theories. To study a particular SCFT in the $\Omega$-background one should promote this five-dimensional solution into a full string or M-theory background. We proceed to provide one such embedding in type IIB supergravity.

\section{Uplift to IIB supergravity}\label{sec:uplift}

The five-dimensional background in (\ref{eq:metCoulomb},\,\ref{eq:XBCoulomb}) can be readily uplifted to a solution of type IIB supergravity using the formulae in \cite{Lu:1999bw}.  The background takes the following simple form in string frame
\begin{eqnarray}\label{eq:IIBbackground}
\dd s^2 &=&  \sqrt{\Delta}\left(\dd s_5^2 + L^2X \dd \theta^2\right) \notag\\
&&+ \f{L^2}{\sqrt{\Delta}}\left(\frac{1}{X}\cos^2\theta\ \dd \Omega_3^2- X^2\sin^2\theta\ \dd\phi^2\right)\,,\notag\\
F_5 &=&  -\f{\im L^4}{g_s}(1+\star_{10})\dd\left(\f{\Delta X^2}{z^4}\right)\w \vol_4\,,\\
B_2 &=&  \im g_s C_2 =  \f{L^2}{4}\e^{-\phi}\sin\theta~B^-\,, ~~~ C_0 =  0\,,~~~ \e^{\Phi} = g_s\,,\notag
\end{eqnarray}
where $g_s$ is the string coupling constant, $\Delta = X\cos^2\theta+X^{-2}\sin^2\theta$, $\dd \Omega_3^2$ is the metric on the round $S^3$, and $\vol_4$ is the volume-form on $\RR^4$ \footnote{It will be interesting to establish whether our solution can arise as a ``near horizon limit'' for the string theory realization of the $\Omega$-deformation discussed in \cite{Hellerman:2011mv} and references thereof.}.

It should be noted that since the five-dimensional supergravity is written in Euclidean signature with an $\SO(1,1)$ gauge group, the internal space in the ten-dimensional IIB background is a deformation of the five-dimensional de Sitter space. This implies that the background above is a solution of Hull's type IIB$^*$ supergravity \cite{Hull:1998vg,Hull:1998ym}. This theory is easily obtained from the standard type IIB supergravity by considering all RR fields to be pure imaginary \footnote{See \cite{Bobev:2018ugk} for a further discussion on this. We use the conventions for type IIB supergravity presented in Appendix A  of \cite{Bobev:2018ugk}.}. Alternatively one can take $\phi \to \im \varphi$ in the background \eqref{eq:IIBbackground} to obtain a purely Euclidean ten-dimensional metric.

When we set $w=0$ the background in \eqref{eq:IIBbackground} is a deformation of Euclidean AdS$_5\times S^5$ and is the supergravity dual of the $\Omega$-deformation of $\mathcal{N}=4$ SYM at the conformal point. The solution preserves 16 supercharges, the ten-dimensional uplift of the 12 spinors in \eqref{eq:5dspinor} and additional 4 spinors similar to $\zeta^{+}$ in \eqref{eq:4dspinors}. Note that the non-trivial fluxes break part of the bosonic symmetries of AdS$_5\times S^5$. For $w\neq 0$ the solution in \eqref{eq:IIBbackground} preserves 12 supercharges and is the holographic dual of the $\Omega$-deformation of the $\SO(4)\times \SO(2)$ invariant ``Coulomb branch'' flow discussed in \cite{Freedman:1999gk}. This flow is triggered by a vev for a scalar operator of conformal dimension 2 in the $\mathbf{20'}$ of the $\SO(6)$ R-symmetry of $\mathcal{N}=4$ SYM. In string theory this vev induces a particular distribution of smeared D3-branes. One can generalize the background in \eqref{eq:IIBbackground} even further by performing an orbifold by a discrete subgroup $\Gamma \subset \SU(2) \subset \SO(4)$ where $\SO(4)$ is the isometry group of $\dd \Omega_3^2$. For the $w=0$ solution this orbifold breaks the 16 supercharges to 12.

The five-dimensional background in (\ref{eq:metCoulomb},\,\ref{eq:XBCoulomb}) can also be uplifted to a solution of eleven-dimensional supergravity using the results in \cite{Gauntlett:2007sm}. This eleven-dimensional solution can be interpreted as a holographic dual to the $\Omega$-deformation of the superconformal theories of ``class $\mathcal{S}$'' studied in \cite{Gaiotto:2009gz}. This will be discussed further in \cite{BGH}. 

\section{Discussion}\label{sec:conclusion}

We presented a supergravity background which describes holographically a four-dimensional $\mathcal{N}=2$ SCFT subject to an $\Omega$-deformation. Given the simple form of this solution it should be possible to use the AdS/CFT dictionary and compute holographically supersymmetric observables in the dual SCFT and compare with results from supersymmetric localizaton. This necessitates a careful study of the large $N$ limit of the Nekrasov partition function for $\mathcal{N}=2$ SCFTs with weakly coupled holographic duals. A natural supersymmetric observable is the free energy of the $\Omega$-deformed SCFT which is given by the on-shell action of the supergravity solution. From the form of the supergravity action in \eqref{eq:Romanslagrangian} and the solution in (\ref{eq:metCoulomb},\,\ref{eq:XBCoulomb}) it appears that the on-shell action does not depend on the deformation parameter $\beta$, which in turn is related to the $\epsilon_{1,2}$ parameters of the $\Omega$-deformation. This should be scrutinized more carefully since there could be finite boundary counterterms which depend on $\beta$ and have to be added in a supersymmetric implementation of the holographic renormalization procedure. Other interesting supersymmetric observables in the $\Omega$-deformed SCFT include line-operators. In the context of the $\mathcal{N}=4$ SYM (and its supersymmetric orbifold extensions) these are described by probe fundamental strings and D1-branes in the type IIB solution \eqref{eq:IIBbackground}. It will be very interesting to compute these holographic observables and understand whether they exhibit non-trivial dependence on the deformation parameter $\beta$. This will be studied in \cite{BGH}.

From the supergravity perspective it is possible to study several generalizations of the background in (\ref{eq:metCoulomb},\,\ref{eq:XBCoulomb},\,\ref{eq:IIBbackground}). One very interesting question is how to deform the IIB supergravity solution in \eqref{eq:IIBbackground} to describe the holographic dual of the $\mathcal{N}=2^{*}$ SYM theory on the $\Omega$-background. It is natural to expect that the superpotential mass parameter in this theory enters non-trivially in the large $N$ limit of the Nekrasov partition function and it should be possible to compute holographically many non-trivial physical observables. We expect that the supergravity solution dual to this more general deformation of $\mathcal{N}=4$ SYM will depend on both $\epsilon_{1,2}$ parameters. It is also possible to combine the $\Omega$-deformation with the Witten-type topological twist, as done in \cite{Nekrasov:2003rj},  and study the holographic dual of this setup. Finally, there is a simple generalization of the solution in (\ref{eq:metCoulomb},\,\ref{eq:XBCoulomb}) to seven-dimensional maximal gauged supergravity which is holographically dual to the $(2,0)$ M5-brane SCFT with an $\Omega$-deformation. We plan to explore these generalizations in \cite{BGH}.

The IIB supergravity solution in \eqref{eq:IIBbackground} is a simple deformation of AdS$_5\times S^5$ which preserves 16 supercharges. Given this high degree of symmetry it is important to understand whether this solution is protected against $\alpha'$ corrections and whether the classical integrability of string theory on AdS$_5\times S^5$ survives the $\Omega$-deformation.
%
%
\section*{Acknowledgements}
We are grateful to F. Benini, S. M. Hosseini, S. Katmadas, J. van Muiden, and A. Van Proeyen  for useful discussions. NB is supported in part by an Odysseus grant G0F9516N from the FWO. FFG is a Postdoctoral Fellow of the Research Foundation - Flanders. KH is supported in part by the Bulgarian NSF grants DN08/3 and N28/5. NB and FFG are also supported by the KU Leuven C1 grant ZKD1118 C16/16/005.

\bibliography{Omega-def}

\end{document}